\documentclass[aps,prb,reprint,showpacs,showkeys,groupedaddress]{revtex4-1}

\usepackage{amsmath, amssymb}
\usepackage{amsmath}
\usepackage{graphicx}
\usepackage{graphics}
\usepackage{subfigure}
\usepackage{color}
\usepackage[colorlinks=true,allcolors=blue]{hyperref}
\usepackage{float}
\newcommand*{\rom}[1]{}

\begin{document}


\title{Hydrodynamic Ratcheting of Tracers in Microchannels: A Detailed Analysis}
\author{Aakash Anand}\thanks{aakash.a@students.iiserpune.ac.in}
\author{A. Bhattacharyay}\thanks{a.bhattacharyay@iiserpune.ac.in}
\affiliation{Department of Physics, Indian Institute of Science Education and Research, Pune, Maharashtra 411008, India.}

\date{\today}

\begin{abstract}
Understanding surface-driven transport is crucial for biological applications and the development of microfluidic devices. In this work, we analyze a local-inversion-symmetry-broken fluid flow model in an undulating microchannel. Surface undulations at a few tens of Hertz in a soft microchannel keep the fluid flow in a low-Reynolds-number regime, enabling the use of a perturbation analysis. Using this, we develop a detailed analysis of the relationship between fluid velocity and tracer particle dynamics, which is crucial for subsequent numerical analysis that shows ratcheting. We show that the tracer particle can undergo ratcheting (which we call hydrodynamic ratcheting) under very specific, physically meaningful conditions. We observe a ratcheting velocity of $\sim 0.15~\mathrm{\mu m/sec}$ for a micrometer-sized particle at room temperature in water when the wavelength of the undulations is of the order of $1~\mathrm{\mu m}$.  

\end{abstract}

\maketitle


\section{INTRODUCTION}
The study of transport phenomena in microchannels and nanochannels is a profoundly influential field of research, with roots extending back several decades. This interest is driven by a dual motivation: first, the fundamental need to understand and replicate the precise transport mechanisms inherent in biological systems, and second, the immense technological promise of harnessing these micro- and nanoscale phenomena to create novel, engineered systems \cite{tagliazucchi2015transport, sirkin2020transport}. The transport phenomena in ion channels of cell membranes are a remarkable exploitation of small-scale cooperation by nature \cite{doyle1998structure}. Biological nanopores, which selectively gate ions and molecules with unparalleled efficiency, have motivated researchers to aim for similar throughput in artificial devices. This ambition to understand the principles at work in biological systems for technological advantages has boosted the development of microfluidics and nanofluidics. As the field grew, a deeper understanding of the underlying physics of fluid flow, based on the principles of classical hydrodynamics with a low Reynolds number \cite{landau1959fluid, batchelor2000introduction, happel1983low, purcell2014life}, became essential. Many detailed studies of phenomena in microchannels and nanochannels have been presented in the literature \cite{stone2004engineering, squires2005microfluidics, whitesides2006origins, karniadakis2006microflows, kirby2010micro, tabeling2023introduction, schoch2008transport}. The precise control that micro- and nanofluidics offers has enabled a wide range of applications, most notably in the separation, sorting, and trapping of particles and cells, which has become a major subdiscipline \cite{sajeesh2014particle, sonker2019separation, yamada2004pinched, di2008equilibrium, gorre1997sorting}.

Microfluidic platforms have played a crucial role in overcoming critical bottlenecks in sample preparation. For example, Mark et al. \cite{mark2010microfluidic} reviewed the use of ``Lab-on-a-chip" systems for techniques such as cell sorting, nucleic acid extraction, and purification to improve sequencing efficiency. In this context, surface-induced flows due to surface corrugations have been explored by Kurzthaler et al. \cite{Kurzthaler_Chase_Stone_2024}, exhibiting helical surface flows.
Jaffrin and Shapiro \cite{annurev:/content/journals/10.1146/annurev.fl.03.010171.000305} have explained in detail the mechanism of peristaltic pumping, which involves a progressive wave of contraction and expansion that moves along a channel, effectively transporting the fluid. Other works focus on cutting-edge applications, such as analyzing the genomes of individual cells, using microfluidic devices to isolate and lyse individual cells, and preparing minute amounts of DNA for sequencing, a process that is otherwise extremely challenging \cite{garcia2019advances}. The physical confinement and geometry of these channels are key to their function. Researchers have analyzed single DNA molecules within nanofluidic channels, utilizing the principle that severe confinement stretches DNA, allowing for detailed analysis without the need for chemical modifications or tying its ends \cite{frykholm2022dna}. Furthermore, geometric features, such as surface undulations, have been shown to alter the particle's diffusion coefficient in a nanochannel, leading to enhanced or reduced transport \cite{marbach2018transport}.

Building on these concepts of geometric influence, our paper develops a detailed and consistent analytical method for the ratcheting of tracer particles \cite{astumian1997thermodynamics, astumian2002brownian, bader1999dna, ethier2018flashing, bier1997brownian} dragged by a local inversion symmetry broken axial velocity field in a microchannel. The velocity profile of the fluid that has been taken into account in the present work corresponds to stationary waves. The fluid front is not traveling, as it generally develops under peristaltic pumping \cite{shapiro1969peristaltic, selverov2001peristaltically, fung1968peristaltic, mishra2003peristaltic, takabatake1988peristaltic, misra2002peristaltic} at the confining boundary of the fluid. In this sense, the present paper addresses a complementary situation to peristaltic flow in microtubes. The main result of this paper quantifies the non-trivial global transport of tracer particles in the presence of a stationary flow field in the fluid, driven by surface undulations. Only the dragging of a tracer by such a flow field cannot result in global (large-scale) transport without being coupled to the Brownian fluctuations of particles, and in the absence of at least a local broken inversion symmetry, which are the basic ingredients for ratcheting. In this context, our paper explores a consistent analytical treatment of the problem and validates the existence of ratcheting transport. The analysis reveals a wealth of information on the details of the interdependence between ratcheting and the geometry, fluid type, and forcing parameters. 

When a tracer particle is subjected to a spatially inverted, symmetry-broken local oscillatory hydrodynamic field, the tracer can experience a drag force that is globally non-directional. Such a hydrodynamic drag force on the tracer generally exists even in the presence of thermal noise. Such a typical situation is amenable to perturbation theory in the low-Reynolds-number regime, where hydrodynamic consistency can be addressed mode by mode in an essentially linear analysis, revealing the interplay between the velocity field and corresponding surface undulations in a very consistent manner. These relationships are essential to understand in order to implement subsequent numerical simulations of tracer particle motion in a microconfined environment. Otherwise, in the absence of precise knowledge of the relationship between the fluid velocity and the surface modes, the boundary conditions on the tracer cannot be implemented, rendering the numerical simulations unreliable. Moreover, striking a balance between diffusion and the local hydrodynamic force on the tracer is crucial for ratcheting that results in global directional transport. The nature of the fluid, its temperature, and the dimensions of the confinement, along with the amplitude and wavelength of the hydrodynamic modes, must cooperate. Such a fine balance of multiple factors cannot be well understood without a detailed analysis of the system's hydrodynamics and its coupling to the tracer. Our paper presents an in-depth account of these ingredients at play in a micro-confined, incompressible, surface-driven fluid.

We organize the paper as follows: first, we present the governing Navier-Stokes equation \cite{kambe2007elementary, kundu1990fluid, landau1959fluid, batchelor2000introduction} and associated boundary conditions for the cylindrical geometry of our model soft channel. Subsequently, we solve the Navier-Stokes equation under the low Reynolds number approximation to find the boundary modes corresponding to the intended velocity field. The velocity field provides the required ratcheting potential for the dynamics of the tracer particle we investigate. Then, we numerically simulate \cite{kloeden2012numerical} the Langevin equation \cite{balakrishnan2008elements, pottier2009nonequilibrium} for the tracer in the presence of the background velocity field, thermal noise, and known dynamics of the confining boundary. We investigate the interplay of various factors, including fluid and drive parameters, to understand and present the ratcheting regime in which the tracer can be driven globally in a given direction. Finally, we discuss our results and conclude the paper.

\section{THEORETICAL MODEL}
\subsection{Fluid Dynamics: Navier-Stokes Equation and Perturbation Analysis}
We aim to derive the fluid velocity field and the associated boundary dynamics in the microchannel. For such flows through microchannels, when surface fluctuations occur at several tens of hertz, the flow is in the low-Reynolds-number regime. Based on this, a perturbation analysis of the flow is developed. The linearity of the problem allows us an analytical understanding.
\begin{figure}[ht]
\includegraphics[width=11.0
cm,height=7.5cm]{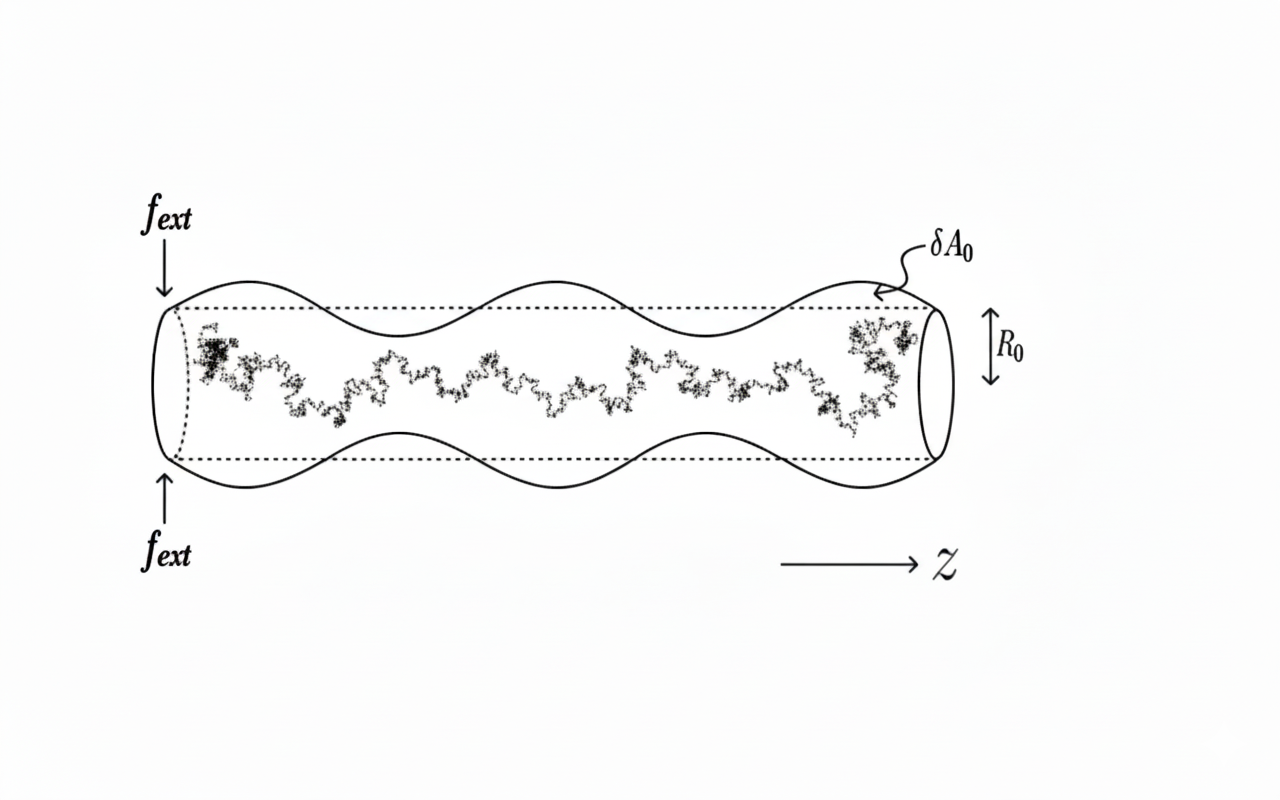}
\caption{The schematic diagram of a microchannel with undulating walls that are circularly symmetric around the $z$-axis. The wall geometry is generated by an external forcing, and a pressure gradient is applied along the tube length.}
\label{fig:1}       
\end{figure}
As shown schematically in Fig.~\ref {fig:1}, the model system we consider is a tracer particle evolving in a cylindrical channel of average radius $R_0$, whose walls, in general, are assumed to support undulations that could be generated by oscillating one of the edges of the tube. We consider the diameter of the tube to be approximately $10~\mu$m.

Navier-Stokes equation for the velocity profile $\boldsymbol{u}(\boldsymbol{r}, t)$ of a fluid flow is
\begin{multline}
    \rho \left(\frac{\partial {\boldsymbol u}}{\partial t}  + (\boldsymbol {u}.\nabla)\boldsymbol {u}\right) = -\nabla P  + \eta \nabla^2 \boldsymbol{u}+\\ \left(\frac{1}{3}\eta + \zeta\right)\nabla(\nabla . {\boldsymbol u}) + \boldsymbol{f}_{\rm ext}
,\label{eq:1}
\end{multline}
supplemented with the continuity equation:
\begin{equation}
    \frac{\partial \rho}{\partial t} + \nabla .(\rho \boldsymbol{u}) = 0.\label{eq:2}
\end{equation}
Where $\rho$ and $\eta$ are, respectively, the mass density and dynamic viscosity of the fluid, and $\zeta$ is the second coefficient of viscosity of the fluid or the bulk viscosity. In Eq.~\ref{eq:1}, $\boldsymbol{f}_{\rm ext}$
 is the body force acting on the fluid (specifically, force per unit volume). 

Throughout this paper, we assume the incompressibility of fluid (i.e., $\frac{d\rho}{d t} = 0$), which gives $\nabla. {\boldsymbol{u}} = 0 $ due to the continuity equation. By this assumption, the third term in the Navier-Stokes equation is left out due to its dependence on $\nabla \cdot \boldsymbol{u}$. Thus, the above equations reduce to the following form:
\begin{equation}
    \rho \left(\frac{\partial \boldsymbol{u}}{\partial t}  + (\boldsymbol{u}.\nabla)\boldsymbol{u}\right) = -\nabla P  + \eta \nabla^2 \boldsymbol{u}+ \boldsymbol{f}_{\rm ext}
,\label{eq:3}
\end{equation}
and
\begin{equation}
    \nabla .{\boldsymbol{u}} = 0.\label{eq:4}
\end{equation} 

We nondimensionalize equations by introducing $r' = r/R_0$, $t' = t/\tau$ ($\tau$ = $2\pi/\omega$), ${\boldsymbol{u'}} = {\boldsymbol{u}}/U$ $ (U = R_0/\tau = \omega R_0/2\pi)$, $P' = P/P_0$ ($P_0 = \eta U/R_0$) as dimensionless variables. With dimensionless quantities, the Eq.~\ref{eq:3} becomes:
\begin{equation}
    \mathrm{Re} \left( \frac{\partial \boldsymbol{u'}}{\partial t'} + (\boldsymbol{u'}.\nabla')\boldsymbol{u'}  \right) = -\nabla' P' +  \nabla'^2{\boldsymbol u'} + \boldsymbol{f}_{\rm ext}'
\label{eq:5},
\end{equation}
where $\mathrm{Re} = \rho U R_0/\eta = \rho \omega R_0^2/2\pi\eta$ is the Reynolds number and $\boldsymbol{f}_{\rm ext}' = \frac{2\pi R_0}{\eta \omega}\boldsymbol{f}_{\rm ext}
$. Typical numerical values considered being $\rho \sim 10^3\;  \text{kg/m}^3$, $\omega \sim 10 - 100 \; \text{rad/s}, R_0 \sim 10 \; \mu\text{m}$, and $\eta \sim 10^{-3} \; \text{Pa s}$, the Reynolds number is approximately $\mathrm{Re} \sim 10^{-4}- 10^{-3}$, which is a very small number.

\par Since we are working in the low Reynolds number regime, $\mathrm{Re}$ serves as a perturbation parameter. Thus, replacing $\mathrm{Re}$ by $\epsilon$ Eq.~\ref{eq:5} takes the shape:
\begin{equation}
     \epsilon \left( \frac{\partial \boldsymbol{u'}}{\partial t'} + (\boldsymbol{u'}.\nabla')\boldsymbol{u'}  \right) = -\nabla' P' +  \nabla'^2{\boldsymbol u'} + \boldsymbol{f}_{\rm ext}'
\label{eq:6}.
\end{equation}
Moreover, the cylindrical geometry of our system, with an undulating surface, makes it convenient to use cylindrical coordinates. In cylindrical coordinates ${\boldsymbol{u}} = v \boldsymbol{\hat{r}} + u \boldsymbol{\hat{z}}$ (assuming axis symmetry). By substituting expressions for Laplacian and convective derivatives in cylindrical coordinates and separating the vector Eq.~\ref{eq:6} into its component equations, we get two equations. The equation for the $\boldsymbol{\hat{z}}$-component is 
\begin{multline}
    \epsilon\left( \frac{\partial u'}{\partial t'} + u' \frac{\partial u'}{\partial z'} + v' \frac{\partial u'}{\partial r'}\right) = -\frac{\partial P'}{\partial z'} +  \frac{\partial^2u'}{\partial z'^2} + \frac{\partial^2 u'}{\partial r'^2} \\+ \frac{1}{r'}\frac{\partial u'}{\partial r'} + f'_z,  \label{eq:7}
\end{multline}
and that in the $\boldsymbol{\hat{r}}$ direction is
\begin{multline}
    \epsilon \left( \frac{\partial v'}{\partial t'} + v' \frac{\partial v'}{\partial r'} + u' \frac{\partial v'}{\partial z'}\right) =  \frac{\partial^2v'}{\partial z'^2} + \frac{\partial^2 v'}{\partial r'^2} + \frac{1}{r'}\frac{\partial v'}{\partial r'} \\ - \frac{v'}{r'^2} + f'_r.\label{eq:8}
\end{multline}

Here, we have used $\boldsymbol{f}_{\rm ext}' = f'_r \boldsymbol{\hat{r}} + f'_z \boldsymbol{\hat{z}}$. This body force must be found out self-consistently because such a force in the bulk of the fluid would eventually be generated.
We expand $u'$ and $v'$ in perturbation series:
\begin{equation}
    u' = u'_0 + \epsilon u'_1 +\epsilon^2 u'_2 + \dots ,  \label{eq:9}
\end{equation}
and
\begin{equation}
     v' = v'_0 + \epsilon v'_1 +\epsilon^2 v'_2 +\dots. \label{eq:10}
\end{equation}

Similarly, the radius $R(z,t)$ and external forcing $\boldsymbol{f}_{\rm ext}
'$ acting on the walls of the microchannel are expanded as:
\begin{equation}
    R(z,t) = R_0 + \epsilon R_1(z,t) + \epsilon^2 R_2(z,t)+\dots , \label{eq:11}
\end{equation}
\begin{equation}
    {f_z}' = \epsilon {f_z}'^{(1)} + \epsilon^2 {f_z}'^{(2)} + \dots , \label{eq:12}
\end{equation}
and
\begin{equation}
    {f_r}' = \epsilon {f_r}'^{(1)} + \epsilon^2 {f_r}'^{(2)} + \dots . \label{eq:13}
\end{equation}
Where $R_0$ is the average radius of the microchannel. $\epsilon R_1$ and $\epsilon^2 R_2$ are, respectively, the first- and second-order undulations present on the surface of the microchannel. As a result of the drive that generates flows in the fluid, the nonlinearity of the flow generates feedback oscillations $\epsilon^2 R_2$ in the second order. This is the scheme we base the perturbation analysis on in a self-consistent manner. The details of the analysis can be found in \cite{anand2025flow}.

\subsubsection{Boundary condition for velocity fields}
To uniquely determine the solution, one needs to impose appropriate boundary conditions on the velocity profile. Boundary conditions describe the fluid behavior at boundaries that the solution must obey at all orders. The velocity profile $\boldsymbol{u}(\boldsymbol{r}, t)$ of the fluid satisfies the following boundary condition (known as the kinematic boundary condition \cite{batchelor2000introduction, kambe2007elementary,shapiro1969peristaltic}):
\begin{multline}
    v(r = R(z,t), z,  t) -  u(r = R(z,t), z, t) \frac{\partial R(z,t)}{\partial z} - \\\frac{\partial R(z,t)}{\partial t} = 0. \label{eq:14}
\end{multline}
Since the elastic tube boundary undulates, a kinematic boundary condition must be adopted, which sets the normal component of the fluid's relative velocity at the boundary to zero, allowing the boundary layers to follow boundary modes.

\subsubsection{Steady flow at Zeroth order}
\par For the sake of completeness, let us first look at the zeroth-order structure of the solution. At zeroth order, the system describes a steady, pressure-driven flow in a rigid, uniform channel. The leading order transport of fluid is due to the pressure gradient. Using Eq.~\ref{eq:7} and \ref{eq:8} one can write the zeroth order equations:
\begin{equation}
     -\frac{\partial P'}{\partial z'} + \frac{\partial^2u'_0}{\partial z'^2} + \frac{\partial^2 u'_0}{\partial r'^2} + \frac{1}{r'}\frac{\partial u'_0}{\partial r'} = 0, \label{eq:15}
\end{equation}
and 
\begin{equation}
     \frac{\partial^2v'_0}{\partial z'^2} + \frac{\partial^2 v'_0}{\partial r'^2} + \frac{1}{r'}\frac{\partial v'_0}{\partial r'} - \frac{v'_0}{r'^2} = 0.\label{eq:16}
\end{equation}

Now, zeroth-order equations solve for steady flow along the $z$-direction in a much simpler setting
\begin{equation}
     \frac{\partial u'_0}{\partial z'} =0\;,\;\frac{\partial^2 u'_0}{\partial z'^2} = 0,\;  \text{and} \; v'_0 = 0.\label{eq:17}
\end{equation}
 Thus, the equation for $u'_0$ becomes: 
 \begin{equation}
     -\frac{\partial P'}{\partial z'} + \frac{\partial^2 u'_0}{\partial r'^2} + \frac{1}{r'}\frac{\partial u'_0}{\partial r'} = 0.\label{eq:18}
\end{equation}
Assuming a constant pressure gradient, solving the equation above yields the following result:
\begin{equation}
    u'_0 = C_1 + \frac{1}{4}\frac{\partial P'}{\partial z'}r'^2, \label{eq:19}
\end{equation}
where $C_1$ is the integration constant. 

Restoring the dimensions of various quantities, we have:
\begin{align}
    u_0 &= U u'_0 \notag\\ & = U \left( C_1 + \frac{1}{4\eta}\frac{\partial P}{\partial z}\frac{R_0}{P_0}r^2\frac{1}{R_0^2} \right) \notag\\
    &= C_2 + \frac{1}{4\eta}\frac{\partial P}{\partial z} r^2. \label{eq:20}
\end{align}
Since $v'_0 = 0$, we have 
\begin{equation}
    v_0 = 0, \label{eq:21}
\end{equation}
where the direction of the longitudinal flow is determined by the pressure head, which could be chosen in either direction in order to make the direction of the bulk flow in a direction opposite to the direction of the ratcheting particles in the fluid.

Note that, instead of imposing the classical no-slip boundary condition on the channel wall, we may consider the presence of a finite slip \cite{lauga2006microfluidics, bocquet2007flow, tretheway2002apparent, neto2005boundary}. This choice is not only physically more realistic for microconfined flows, but it will also remove divergence ($t$ dependent growth in the expression of $R_1(z,t)$) that would otherwise arise in the boundary profile. Accordingly, the velocity field is expressed as the sum of two contributions: the first part corresponds to the usual Poiseuille component, which vanishes at the wall, while the second part accounts for the finite slip velocity $v_{\rm slip}$ that remains nonzero at the boundary. Then,
\begin{equation}
    u_0 = -\frac{1}{4\eta}\frac{\partial P}{\partial z}\left(R_0^2 - r^2\right) + v_{\rm slip},\label{eq:22}
\end{equation}
where the continuity equation at zeroth order is
\begin{equation}
     \frac{1}{r}\frac{\partial}{\partial r}(rv_0) + \frac{\partial u_0} {\partial z} = 0.\label{eq:23}
\end{equation}
With these expressions for $u_0$ and $v_0$, the zeroth-order continuity equation trivially satisfies, as the pressure gradient $\partial P/\partial z$, which is a constant \cite{anand2025flow}.

\subsubsection{First order equations}
The first order equations obtained from Eq.~\ref{eq:7} and \ref{eq:8} are:
\begin{equation}
     \frac{\partial^2 u'_1}{\partial z'^2} + \frac{\partial^2 u'_1}{\partial r'^2} + \frac{1}{r'}\frac{\partial u'_1}{\partial r'} = -{f_z}'^{(1)},\label{eq:24}
\end{equation}
and
\begin{equation}
     \frac{\partial^2 v'_1}{\partial z'^2} + \frac{\partial^2 v'_1}{\partial r'^2} + \frac{1}{r'}\frac{\partial v'_1}{\partial r'} - \frac{v'_1}{r'^2} = -{f_r}'^{(1)},\label{eq:25}
\end{equation}
where ${f_z}'^{(1)}$ and ${f_r}'^{(1)}$, respectively, represent the components $z$ and $r$ of the dimensionless forces (per unit volume) experienced by the fluid due to the surface undulations induced velocity field present in the microchannel. 

Instead of defining the forces and solving for the velocity, we adopt the inverse approach.  We prescribe a desired time-dependent velocity profile $u'_1$ that can act as a ratcheting potential for a tracer particle and then use Eq.~\ref{eq:24} to determine the corresponding forces required to generate it in a self-consistent manner. To act as a ratcheting potential, we must choose a velocity field $u_1'$ to be of a spatially inversion-symmetry broken form modulated by temporal undulations, which will act as a non-equilibrium force to the tracer particle. We do this by using the minimal coupling of the first two spatial harmonics of a sinusoidal wave, which leads to the following form:
\begin{equation}
\begin{aligned}
u'_1(r',z',t') &= C'_1 \, \sin(\omega' t' + \phi) \, \cos(\omega' t') \\
&\quad \times \Bigl( \cos(2k'z') - 2 \, \cos(k'z') \Bigr),
\end{aligned}
\label{eq:26}
\end{equation}
where $k' = kR_0$ and  $\omega' = 2\pi$ as usual. Note that the linearity of Eq.~\ref{eq:24} and \ref {eq:25} allows us to consider the velocity profile as a superposition of modes, and here lies the merit of the analytical method that we have developed.

As the continuity equation at first order is of the following form:
\begin{equation}
    \frac{1}{r'}\frac{\partial}{\partial r'}(r'v_1') + \frac{\partial u_1'}{\partial z'} = 0, \label{eq:27} 
\end{equation}
by using the expression for $u'_1$, we can integrate the continuity equation to find the corresponding radial velocity component $v'_1$, which is given by
\begin{equation}
\begin{aligned}
v_1' &= C_1' \, k' \, r' \, \sin(\omega' t' + \phi) \, \cos(\omega' t') \\
&\quad \times \Bigl( \sin(2k'z') - \sin(k'z') \Bigr).
\end{aligned}
\label{eq:28}
\end{equation}
Restoring the dimensions and absorbing the dimensionless constant $C'_1$ in the characteristic velocity scale $U = \omega R_0/2\pi$, we get the following expressions for the first-order velocity fields.
\begin{equation}
\begin{aligned}
u_1(r,z,t) &= U \, \sin(\omega t + \phi) \, \cos(\omega t) \\
&\quad \times \Bigl( \cos(2kz) - 2\cos(kz) \Bigr),
\end{aligned}
\label{eq:29}
\end{equation}
and 
\begin{equation}
\begin{aligned}
v_1(r,z,t) &= k r \, U \, \sin(\omega t + \phi) \, \cos(\omega t) \\
&\quad \times \Bigl( \sin(2kz) - \sin(kz) \Bigr).
\end{aligned}
\label{eq:30}
\end{equation}

The complete velocity fields in the channel are given by
\begin{multline}
u(r,z,t) = -\frac{1}{4\eta}\frac{\partial P}{\partial z}(R_0^2 - r^2) + v_{\rm slip} + \\
{\rm Re} \frac{\omega R_0}{2\pi} \, \sin(\omega t + \phi) \, \cos(\omega t)  \Bigl( \cos(2kz) - 2\cos(kz) \Bigr),\label{eq:31}
\end{multline}
and 
\begin{multline}
v(r,z,t) = {\rm Re}\; k r \, \frac{\omega R_0}{2\pi} \, \sin(\omega t + \phi) \, \cos(\omega t) \\
\times \Bigl( \sin(2kz) - \sin(kz) \Bigr),
\label{eq:32}
\end{multline}
where, ${\rm Re} = \rho \omega R_0^2/2\pi\eta$.
To self consistently determine the forces ${f_z}'^{(1)}$ and ${f_r}'^{(1)}$ we substitute the expression of $u_1'$ and $v_1'$ in Eq.~\ref{eq:24} and \ref{eq:25} respectively and we get:
\begin{multline}
    {f_z}'^{(1)} 
    = 2C_1' k'^2 
      \sin(\omega' t' + \phi)\cos(\omega' t') \\
      \times \Bigl[\, 2\cos(2k'z') - \cos(k'z') \,\Bigr], \label{eq:33}
\end{multline}
and
\begin{multline}
    {f_r}'^{(1)} 
    = C_1' k'^3 r' 
      \sin(\omega' t' + \phi)\cos(\omega' t') \\
      \times \Bigl[\, 4\sin(2k'z') - \sin(k'z') \,\Bigr], \label{eq:34}
\end{multline}
which are quite regular and devoid of singularity.

Restoring the dimensions in the above equations gives:
\begin{multline}
    f_z^{(1)} 
    = 2\,\eta\, U k^2 \,
      \sin(\omega t + \phi)\cos(\omega t) \\
      \times \Bigl[\, 2\cos(2kz) - \cos(kz) \,\Bigr], \label{eq:35}
\end{multline}
and
\begin{multline}
    f_r^{(1)} 
    = \eta\, U k^3 r \,
      \sin(\omega t + \phi)\cos(\omega t) \\
      \times \Bigl[\, 4\sin(2kz) - \sin(kz) \,\Bigr].\label{eq:36}
\end{multline}
It should be noted that the first order velocities $u_1$ and $v_1$ together with the above determined forces $f_z^{(1)} $, $f_r^{(1)} $ together with $R_1(z,t)$ (determined in the next section) will close the problem in a consistent manner.

\subsubsection{First-order Radius Correction}

The first-order kinematic boundary condition is as follows:
\begin{equation}
\begin{aligned}
&v_1(r = R_0, z, t) 
- u_0(r = R_0, z, t) \frac{\partial R_1(z, t)}{\partial z} \\
&\quad - u_1(r = R_0, z, t) \frac{\partial R_0}{\partial z} 
- \frac{\partial R_1(z, t)}{\partial t} = 0.
\end{aligned}
\label{eq:37}
\end{equation}
Since $R_0$ is constant, the third term in the above equation is zero. Substituting the expression of various quantities, we get the following equation:
\begin{equation}
\begin{aligned}
v_{\rm slip} \frac{\partial R_1(z,t)}{\partial z} 
&+ \frac{\partial R_1(z,t)}{\partial t} \\
&= k R_0 U \, \sin(\omega t + \phi) \\
&\quad \times \cos(\omega t) \, 
    \Bigl( \sin(2k z) - \sin(k z) \Bigr).
\end{aligned}
\label{eq:38}
\end{equation}
This is a first-order linear partial differential equation in $R_1(z,t)$. 

To solve this PDE, we use the method of characteristics \cite{asmar2016partial, myint2007linear}. The characteristic equations are given by:
\begin{align}
    \frac{dz}{d \tau} &= v_{\rm slip}, \label{eq:39} \\
    \intertext{and}
    \frac{dt}{d \tau} &= 1. \label{eq:40}
\end{align}
The solutions of the above equations are as follows:
\begin{align}
    z(\tau) &= v_{\text{slip}}\,\tau + C_1, \label{eq:41} \\    \intertext{and}
    t(\tau) &= \tau + C_2. \label{eq:42}
\end{align}
Thus, the characteristic line is given by:
\begin{equation}
    z - v_{\rm slip}t = C ( = C_1 - v_{\rm slip} C_2).\label{eq:43} 
\end{equation}

Along these characteristic lines, the partial differential equation reduces to an ordinary differential equation for $R_1$. For a particular characteristic line obtained by $C = \zeta$ (i.e., $z  = v_{\rm slip}t + \zeta$) we get the following ordinary differential equation for $R_1$:
\begin{multline}
    \frac{d R_1}{dt} = \beta \sin(\omega t + \phi)\cos(\omega t) \\ \times \Bigl[\sin(2k(v_{\rm slip}t + \zeta)) - \sin(k(v_{\rm slip}t + \zeta))\Bigr], \label{eq:44} 
\end{multline}
where $\beta = kUR_0$ and the solution to Eq.~\ref{eq:44} is:
\begin{multline}
    R_1(v_{\rm slip}t + \zeta, t) = \beta \int_0^t \; d\tau \sin(\omega \tau + \phi) \cos(\omega \tau) \\ \times \Bigl[\sin(2k(v_{\rm slip}\tau + \zeta)) - \sin(k(v_{\rm slip}\tau + \zeta))\Bigr].\label{eq:45} 
\end{multline}
Where $\tau$ represents the dummy variable of integration. Now, replacing $\zeta$ by $z - v_{\rm slip}t$, we get:
\begin{multline}
R_1(z, t) = \beta \int_0^t d\tau \; 
    \sin(\omega \tau + \phi) \, \cos(\omega \tau)  \\
 \times \Bigl[ 
    \sin\bigl(2k(z - v_{\rm slip}(t - \tau))\bigr) 
    -  \sin\bigl(k(z - v_{\rm slip}(t - \tau))\bigr) 
\Bigr]. \label{eq:46}
\end{multline}
Using the identity,
\begin{equation}
    \sin A \cos B = \frac{1}{2}(\sin(A+B) + \sin(A-B)), \label{eq:47} 
\end{equation}
we get
\begin{multline}
    R_1(z, t) = \frac{\beta}{2} \int_0^t d\tau \; 
    \Bigl( \sin \phi + \sin(2\omega \tau + \phi) \Bigr) \\
    \times \Bigl[ \sin(a_2 + b_1 \tau) - \sin(a_1 + b_1 \tau) \Bigr]. \label{eq:48} 
\end{multline}
Where 
\begin{align}
    a_2 &= 2kz - 2k v_{\text{slip}} t, \label{eq:49} \\
    b_2 &= 2k v_{\text{slip}}, \label{eq:50} \\
    a_1 &= kz - k v_{\text{slip}} t, \label{eq:51} \\
    \text{and} \qquad b_1 &= k v_{\text{slip}}. \label{eq:52}
\end{align}

By defining
\begin{equation}
    S(t;a,b) = \int_0^t  \sin(a+b \tau)\; d\tau , \label{eq:53} 
\end{equation}
and 
\begin{equation}
    K(t; a, b) = \int_0^t \sin(2\omega  \tau + \phi) \sin(a+b \tau)\;d\tau , \label{eq:54} 
\end{equation}
we get the following
\begin{multline}
    R_1(z,t) = \frac{\beta}{2} \Bigl\{ 
        \sin \phi \, \Bigl[ S(t; a_2, b_2) - S(t; a_1, b_1) \Bigr] \\
        + \Bigl[ K(t; a_2, b_2) - K(t; a_1, b_1) \Bigr] 
    \Bigr\}. \label{eq:55} 
\end{multline}
Where the S-integral is
\begin{equation}
    S(t;a,b) = \frac{\cos a  - \cos(a+bt)}{b}. \label{eq:56} 
\end{equation}
And the K-integral reads
\begin{multline}
    K(t; a,b) = \frac{1}{2} \int_0^t 
    \Bigl( \cos\bigl((2\omega - b)\tau + \phi - a \bigr) \\
    - \cos\bigl((2\omega + b)\tau + \phi + a \bigr) \Bigr) 
    \, d\tau. \label{eq:57} 
\end{multline}
Solving the above gives:
\begin{multline}
    K(t;a,b) = \frac{1}{2} \Biggl( 
        \frac{ \sin\bigl((2\omega - b)t + \phi - a \bigr)  - \sin(\phi -a)}{2\omega - b} \\
        \quad - \frac{ \sin\bigl((2\omega + b)t + \phi + a \bigr) 
        - \sin(\phi + a) }{2\omega + b} 
    \Biggr). \label{eq:58} 
\end{multline}

Writing $\Omega = k v_{\rm slip}$, we get the following analytical expression for $R_1(z,t)$.
\begin{equation}
\begin{aligned}
R_1(z,t) &= \frac{\beta}{2} \Bigl\{ 
    \sin \phi \,
    \frac{\cos\bigl(2kz - 2\Omega t\bigr) - \cos(2kz)}{2 \, \Omega} \\
&\quad - \sin \phi \,
    \frac{\cos\bigl(kz - 2\Omega t\bigr) - \cos(kz)}{\Omega} \\
&\quad + \frac{1}{2} \Bigl[
    \frac{\sin(2\omega t - 2kz + \phi) + \sin(2kz - 2 \Omega t - \phi)}{2\omega - 2 \Omega} \\
&\quad + \frac{\sin(2\omega t + 2kz + \phi) - \sin(2kz - 2 \Omega t + \phi)}{2\omega + 2 \Omega} \\
&\quad + \frac{\sin(2\omega t - kz + \phi) + \sin(kz - \Omega t - \phi)}{2\omega - \Omega} \\
&\quad + \frac{\sin(2\omega t + kz + \phi) - \sin(kz - \Omega t + \phi)}{2\omega + \Omega}
    \Bigr]
\Bigr\}.
\end{aligned}
\label{eq:59}
\end{equation}
\par One should note that the model yields an unphysical result in the limit of zero slip velocity. Specifically, as $v_{\rm slip} \to 0$, the expression for $R_1(z,t)$ shows unbounded temporal growth. Consequently, the existence of a finite slip velocity is a key physical necessity in such situations. The $v_{\rm slip}$ will be picked by the system depending upon the amplitude of the surface undulations excited to sustain a stationary amplitude of the flow.

After rearranging the above equation and substituting $\beta = kUR_0$ we obtain:
\begin{equation}
\begin{aligned}
R_1(z,t) &= \frac{kUR_0}{2 \Omega} \Bigl\{ 
    \sin \phi \,
    \frac{\cos\bigl(2kz - 2\Omega t\bigr) - \cos(2kz)}{2} \\
&\quad - \sin \phi \,
    \frac{\cos\bigl(kz - 2\Omega t\bigr) - \cos(kz)}{1} \\
&\quad + \frac{1}{2} \Bigl[
    \frac{\sin(2\omega t - 2kz + \phi) + \sin(2kz - 2 \Omega t - \phi)}{2(\omega/\Omega - 1)} \\
&\quad + \frac{\sin(2\omega t + 2kz + \phi) - \sin(2kz - 2 \Omega t + \phi)}{2(\omega/\Omega + 1)} \\
&\quad + \frac{\sin(2\omega t - kz + \phi) + \sin(kz - \Omega t - \phi)}{2(\omega/\Omega - 1/2)} \\
&\quad + \frac{\sin(2\omega t + kz + \phi) - \sin(kz - \Omega t + \phi)}{2(\omega/\Omega+ 1/2)}
    \Bigr]
\Bigr\}.
\end{aligned}
\label{eq:60}
\end{equation}
Then substituting $U = \omega R_0/2\pi$ we get,
\begin{equation}
\begin{aligned}
R_1(z,t) &= \frac{\omega R_0^2}{4 \pi v_{\rm slip}} \Bigl\{ 
    \frac{1}{2} \sin \phi \,
    \bigl(\cos\bigl(2kz - 2\Omega t\bigr) - \cos(2kz)\bigr) \\
&\quad - \sin \phi \,
    \bigl(\cos\bigl(kz - 2\Omega t\bigr) - \cos(kz)\bigr) \\
&\quad + \frac{1}{4} \Bigl[
    \frac{\sin(2\omega t - 2kz + \phi) + \sin(2kz - 2 \Omega t - \phi)}{(\omega/\Omega - 1)} \\
&\quad + \frac{\sin(2\omega t + 2kz + \phi) - \sin(2kz - 2 \Omega t + \phi)}{(\omega/\Omega + 1)} \\
&\quad + \frac{\sin(2\omega t - kz + \phi) + \sin(kz - \Omega t - \phi)}{(\omega/\Omega - 1/2)} \\
&\quad + \frac{\sin(2\omega t + kz + \phi) - \sin(kz - \Omega t + \phi)}{(\omega/\Omega+ 1/2)}
    \Bigr]
\Bigr\}.
\end{aligned}
\label{eq:61}
\end{equation}

Thus, the boundary profile $R(z,t)$ is given by:
\begin{equation}
    R(z,t) = R_0 + {\rm Re}\; R_1(z,t), \label{eq:62}
\end{equation}
where $R_1(z,t)$ is given by the Eq.~\ref{eq:61}. 
From above, the amplitude of surface undulations $\delta A_0$ can be read off as $\delta A_0 = {\rm Re}\; \omega R_0^2/4\pi v_{\rm slip}$. Thus,
\begin{equation}
    \delta A_0 = \frac{\rho \omega^2 R_0^4}{8 \pi^2 \eta v_{\rm slip}}.\label{eq:63}
\end{equation}

One may rewrite $R_1(z,t)$ in the form
\begin{multline}
R_1(z,t) = \frac{\beta}{2} \Big[ 
\cos(2kz)\, G_1(t) + \sin(2kz)\, G_2(t) 
+ \\ \cos(kz)\, H_1(t) + \sin(kz)\, H_2(t) ]\Big., \label{eq:64} 
\end{multline}
where
\begin{multline}
G_1(t)=\sin\phi\,\frac{\cos(2\Omega t)-1}{2\Omega} \\[6pt]
+\frac{1}{2}\Bigg[
\frac{\sin(2\omega t+\phi)-\sin(2\Omega t+\phi)}{2\omega-2\Omega} \\[6pt]
+\frac{\sin(2\omega t+\phi)+\sin(2\Omega t-\phi)}{2\omega+2\Omega}
\Bigg],  \label{eq: 65}
\end{multline}
\begin{multline}
G_2(t)=\sin\phi\,\frac{\sin(2\Omega t)}{2\Omega} \\[6pt]
+\frac{1}{2}\Bigg[\frac{\cos(2\Omega t+\phi)-\cos(2\omega t+\phi)}{2\omega-2\Omega} \\[6pt]
+\frac{\cos(2\omega t+\phi)-\cos(2\Omega t-\phi)}{2\omega+2\Omega}\Bigg], \label{eq: 66}
\end{multline}
\begin{multline}
H_1(t)=-\sin\phi\,\frac{\cos(\Omega t)-1}{\Omega} \\[6pt]
+\frac{1}{2}\Bigg[\frac{\sin(2\omega t+\phi)-\sin(\Omega t+\phi)}{2\omega-\Omega} \\[6pt]
+\frac{\sin(2\omega t+\phi)+\sin(\Omega t-\phi)}{2\omega+\Omega}\Bigg], \label{eq: 67}
\end{multline}
\begin{multline}
H_2(t)=-\sin\phi\,\frac{\sin(\Omega t)}{\Omega} \\[4pt]
+\frac{1}{2}\Bigg[\frac{\cos(\Omega t+\phi)-\cos(2\omega t+\phi)}{2\omega-\Omega} \\[6pt]
+\frac{\cos(2\omega t+\phi)-\cos(\Omega t-\phi)}{2\omega+\Omega}\Bigg]. \label{eq: 68}
\end{multline}

Note that Eqs.~\ref{eq:64}-\ref{eq: 68} indicate the presence of stationary wave profiles on the surface, giving rise to the intended flow structures in the bulk. In each of the temporal oscillation terms, there is a temporal oscillation frequency of $2\omega$ present in each term, and on top of that, there is an oscillation frequency $\Omega$ present, which is picked up by the system depending. Obviously, when generating such stationary surface modes, one does not have complete control over exciting traveling modes from either side of the tube to result in the intended surface profile. The system will determine a part of it by the selection of the slip velocity.

\subsection{Langevin Model for Tracer Dynamics}

\subsubsection{Overdamped Langevin Equation}
To model the dynamics of a microscopic tracer particle suspended in the fluid, we employ the Langevin framework. This approach treats the particle's motion as a combination of deterministic drag from the surrounding fluid and stochastic kicks from thermal fluctuations. The equation of motion is given by:
\begin{multline}
    m\frac{d {\boldsymbol u}_p(t)}{dt} = -\gamma({\boldsymbol u}_p(t) - {\boldsymbol u}({ \boldsymbol r_p}(t), t)) + \sqrt{2\gamma k_B T}\boldsymbol{\xi}(t).\label{eq:69} 
\end{multline}

In the above equation, $\gamma$ is the damping constant of the tracer particle in the fluid, $T$ is the temperature, and $k_B$ is the Boltzmann constant. Here $\boldsymbol{\xi}(t) = (\xi_r(t), \xi_{\theta}(t), \xi_z(t))$ represents the noise vector whose components are delta-correlated stationary Gaussian processes with zero mean, satisfying:
\begin{equation}
   \langle \boldsymbol{\xi}(t) \rangle = 0 ,\label{eq:70} 
\end{equation}
and
\begin{equation}
    \langle \xi_i(t)\,\xi_j(t') \rangle = \,\delta_{ij}\,\delta(t - t'),
    \qquad i,j \in \{r, \theta, z\}. \label{eq:71} 
\end{equation}

Here, ${\boldsymbol u}_p(t)$ is the velocity of the tracer and ${\boldsymbol u}({ \boldsymbol r}_p(t), t)$ denotes the velocity of fluid flow being evaluated at the position of the tracer particle $ { \boldsymbol r}_p(t)$.
Since we are working in the low Reynolds number regime, we take the overdamped limit of the Langevin equation, in which the inertial term on the left-hand side vanishes, reducing the equation to:
\begin{equation}
    0 = -\gamma({\boldsymbol u}_p(t) - {\boldsymbol u}({ \boldsymbol r}_p(t), t)) + \sqrt{2\gamma k_B T}\boldsymbol{\xi}(t). \label{eq:72} 
\end{equation}

We can identify the particle's translational diffusion coefficient, $D_0$, through the Stokes-Einstein relation as $D_0 = k_B T / \gamma$. The use of this relation is justified because ratcheting in a low Reynolds number fluid flow is a weakly non-equilibrium phenomenon. Thus, the Eq.~\ref{eq:72} can then be written as:
\begin{equation}
   \frac{d {\boldsymbol r}_p(t)}{dt} =  {\boldsymbol u}({ \boldsymbol r}_p(t), t)) + \sqrt{2D_0}\boldsymbol{\xi}(t). \label{eq:73} 
\end{equation}
Writing down the above vector equation in component cylindrical coordinates, we get:
\begin{align}
\frac{d z_p(t)}{dt} &= u_{z} \big(r_p(t), z_p(t), t\big) + \sqrt{2D_0}\,\xi_z(t), \label{eq:74} \\
\frac{d r_p(t)}{dt} &= u_{r} \big(r_p(t), z_p(t), t\big) + \frac{D_0}{ r_p} + \sqrt{2D_0}\,\xi_r(t), \label{eq:75} \\
\frac{d \theta_p(t)}{dt} &= \sqrt{\frac{2D_0}{r_p^2}}\,\xi_{\theta}(t). \label{eq:76}
\end{align}

Note that the radial equation Eq.~\ref{eq:75} includes a term $D_0/r_p$ (where $D_0=k_B T / \gamma$). This is not an external potential, but a known geometric correction that appears when interpreting a stochastic differential equation \cite{oksendal2013stochastic, thygesen2023stochastic} in curvilinear coordinates. This is known as the geometric It\^o correction \cite{risken1989fokker, gardiner2009springer}.

\subsubsection{Boundary condition for tracer particle}
For a tracer particle confined within an impermeable channel, a reflecting boundary condition is imposed on the walls. This physically ensures that no particle can cross the boundary, and thus the total probability $P({\boldsymbol r},t)$ of finding the particle inside the channel is conserved.
Mathematically, this is expressed by stating that the component of the probability current ${\boldsymbol J}({\boldsymbol r}, t)$, that is normal (perpendicular) to the boundary surface, must be zero.
\begin{equation}
    \boldsymbol{J}(\boldsymbol{r}, t) \cdot \boldsymbol{n} = 0 \quad \text{at the boundary } r = R(z,t),
    \label{eq:77}
\end{equation}
where $\boldsymbol{n}$ is the unit vector normal to the channel wall. Substituting the definition of the probability current, $\boldsymbol{J} = \boldsymbol{u}P - D_0\nabla P$, gives the full form of the boundary condition:
\begin{align}
\left[ \boldsymbol{u}(\boldsymbol{r},t) P(\boldsymbol{r}, t) - D_0 \nabla P(\boldsymbol{r}, t) \right] \cdot \boldsymbol{n} &= 0.
\label{eq:78}
\end{align}

\subsubsection{Equations of Motion for Simulations}

The dynamics of the particle are given by:
\begin{align}
\frac{dz_p}{dt} &= u_0\big(r_p(t),z_p(t),t\big) + \mathrm{Re}\;u_1\big(r_p(t),z_p(t),t\big) \nonumber \\
&\qquad + \sqrt{2D_0}\,\xi_z(t), \label{eq:79} \\
\frac{dr_p}{dt} &= \mathrm{Re}\;v_1\big(r_p(t),z_p(t),t\big) + \frac{D_0}{r_p(t)} + \sqrt{2D_0}\,\xi_r(t), \label{eq:80} \\[6pt]
\frac{d\theta_p}{dt} &= \sqrt{\frac{2D_0}{\,r_p(t)^2}}\,\xi_\theta(t). \label{eq:81}
\end{align}

Now, substituting the expression of various terms, we get the following equations:
\begin{align}
\frac{dz_p}{dt} &= -\frac{1}{4\eta}\,\frac{\partial P}{\partial z}\,(R_0^2 - r_p^2) + v_{\rm slip} \nonumber \\
&\quad + \mathrm{Re}\frac{\omega R_0}{2\pi}\sin(\omega t+\phi)\cos(\omega t) \nonumber \\
&\quad \times \big[\cos(2kz_p) - 2\cos(kz_p)\big] + \sqrt{2D_0}\,\xi_z(t),\label{eq:82}
\end{align}
\begin{align}
\frac{dr_p}{dt} &= \mathrm{Re}\; kr_p \frac{\omega R_0}{2\pi} \sin(\omega t + \phi) \cos(\omega t) \nonumber \\
&\quad \times \big[ \sin(2kz_p) - \sin(kz_p) \big] + \frac{D_0}{r_p} +\sqrt{2D_0}\, \xi_r(t), \label{eq:83} \\
\frac{d\theta_p}{dt} &= \sqrt{\frac{2D_0}{r_p^2}} \, \xi_\theta(t). \label{eq:84}
\end{align}
Equations~\ref {eq:82}, \ref{eq:83}, and \ref{eq:84} are our working equations for numerical simulations.

\section{NUMERICAL RESULTS}

To investigate the transport properties of the tracer particle under the influence of the undulating walls, we performed numerical simulations \cite{kloeden2012numerical} of the overdamped Langevin equations. The fluid velocity field in general includes zeroth-order and first-order contributions as given in Eq.~\ref{eq:31} and \ref {eq:32}. To illustrate the effect of hydrodynamic ratcheting, we have considered only the first-order correction to the velocity fields, which simulates the dynamics of the tracer particle. This is achieved by applying a purely reflecting boundary condition, which confines the tracer particle's position to the walls of the channel. 

It should be noted that in this paper, we have considered purely reflecting boundary conditions for the tracer particle, whereas in real situations, the boundary condition can be more complex, including partially absorbing and partially reflecting types, depending on the channel wall structure. But for the sake of simplicity, at the first stage of reporting the procedure, we have considered here only the purely reflecting boundary condition. The sensitivity of ratcheting to other boundary effects is intended to be a subject matter of future investigations on our part.

\begin{figure}[htbp]
    \centering
    \subfigure[]{\label{fig:z_vs_t}\includegraphics[width=0.32\textwidth]{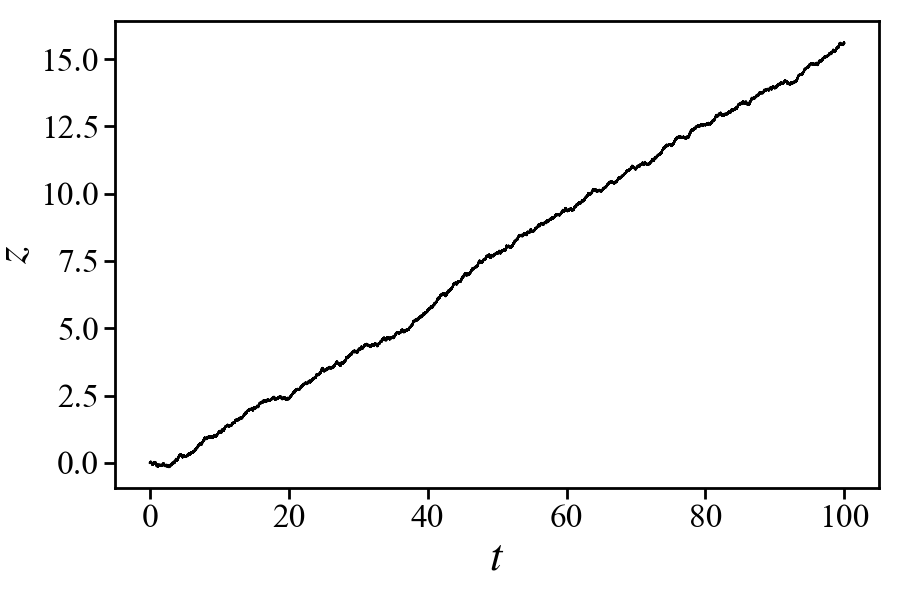}} 
    \subfigure[]{\label{fig:r_vs_t}\includegraphics[width=0.32\textwidth]{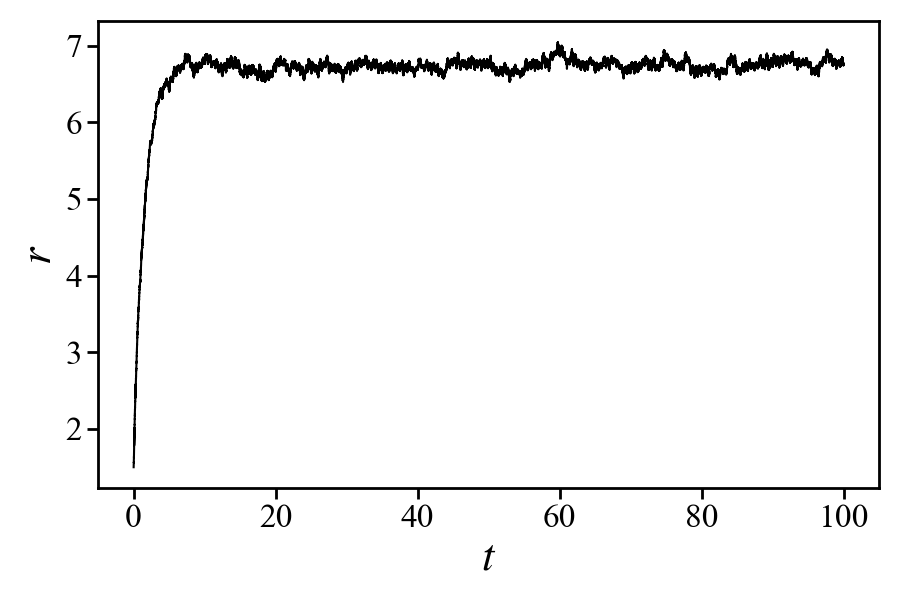}} 
    \subfigure[]{\label{fig:theta_vs_t}\includegraphics[width=0.32\textwidth]{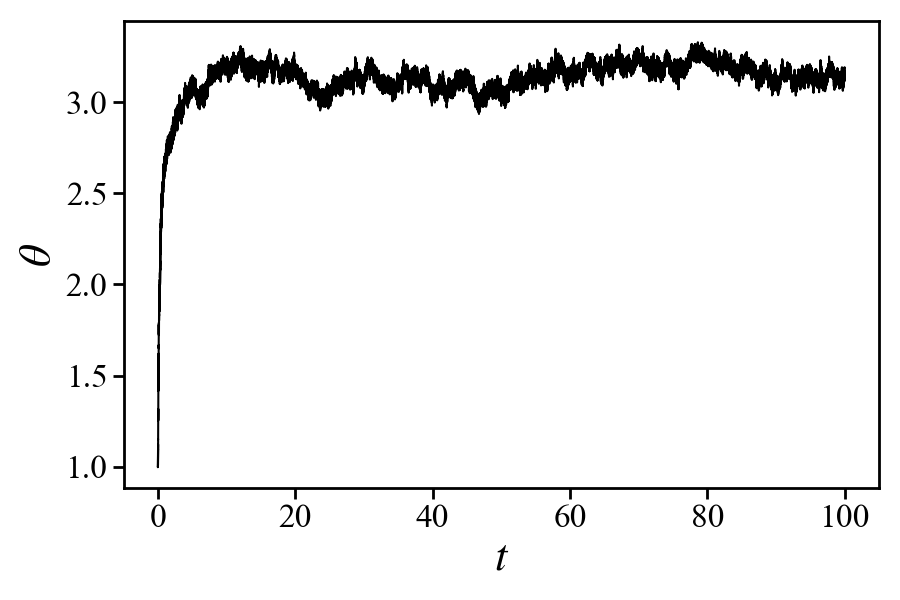}}
    
    \caption{Trajectory of a tracer particle in cylindrical coordinates. The simulation parameters used are: $\omega = 70~\text{rad s}^{-1}, \lambda = 1.0~\mu\text{m}, R_0 = 10~\mu\text{m}, D_0 = 1.0~\mu\text{m}^2\text{ s}^{-1}, v_{\text{slip}} = 1.0~\mu\text{m s}^{-1}, \phi = -\pi/2~\text{rad}.$
    (a) Axial coordinate $z$ (in $\mu$m) as a function of time $t$ (in sec), showing the net drift of the particle. 
    (b) Radial coordinate $r$ (in $\mu$m) as a function of time $t$ (in sec). 
    (c) Angular coordinate $\theta$ (in radians) as a function of time $t$ (in sec).}
    \label{fig:2}
\end{figure}

Fig.~\ref{fig:2} illustrates the ratcheting dynamics, simulated using Eq.~\ref {eq:82}, \ref {eq:83}, and \ref {eq:84} in the absence of a pressure head or slip velocity. This simplification is justified because the pressure-driven field can be minimized or reversed by adjusting the pressure gradient, and this global field can drag a tracer along, resulting in a trivial outcome. The ratcheting gives rise to a global directional velocity of the tracer, driven by the undulating local field in the presence of noise. This result is non-trivial because, otherwise, the undulating local field would not lead to large-scale directed transport. 

The simulation time step is set to $\Delta t = 10^{-4}\,\text{s}$, and we employ the Euler-Maruyama method for numerical integration. We have calculated the tracer particle's velocity by averaging over 1000 realizations. In Fig.~\ref{fig:z_vs_t}, we have plotted the evolution of the longitudinal coordinate $z$ of the tracer particle. It clearly shows the main result of our work that the inversion-symmetry-broken local velocity undulations generate a ratcheting mechanism. The velocity obtained from the simulation is approximately 0.15 $\mu \text{m/s}$. 

The next Fig.~\ref{fig:r_vs_t} shows the evolution of the radial ($r$) coordinate. It can be seen that the particle's radial coordinate moves away from the channel axis. This is what is expected from Eq.~\ref{eq:83} for a cylindrical channel. The term $1/r$ acts as a negative logarithmic potential ($\propto -\ln r$) for the evolution of the radial coordinate of the tracer particle, pushing the particle away from the axis of the channel. Similarly, the evolution of the angular coordinate ($\theta$) with time is shown in Fig.~\ref{fig:theta_vs_t}. The time evolution of the ensemble-averaged angular coordinate shows a saturation at long times. However, it is essential to note that this behavior does not indicate the true confinement of the angular variable. The angular dynamics are governed by a pure diffusive process, with the effective diffusion coefficient determined by the tracer particle's radial coordinate. This radial dependence reduces the rate at which the angular coordinate spreads over the whole angular range. Consequently, upon averaging over a large number of realizations (here, 1000 ensembles), the symmetric diffusive spreading around the initial angle leads to an apparent saturation of the mean angular displacement. This apparent confinement is therefore a statistical artifact of ensemble averaging and slow diffusion, rather than a physical localization mechanism. The parameter values used in the simulations are $\omega = 70 \;\text{rad s}^{-1}$, $\lambda = 1.0 \;\mu$m, $R_0 = 10 \;\mu$m, $D_0 = 1.0 \;\mu\text{m}^2\,\text{s}^{-1}$, $v_{\rm slip} = 1.0 \;\mu\text{m}\,\text{s}^{-1}$ and $\phi = -\pi/2$ rad. 

The drag coefficient (or damping coefficient) $\gamma$ for the dynamics of a tracer particle in the fluid can be written in terms of the radius $a$ of the tracer particle, the dynamic viscosity $\eta$ of the fluid, by the Stokes formula: $\gamma = 6\pi \eta a$. The value of $D_0 = 1.0 \,\mu\text{m}^2\text{s}^{-1}$ can be used to interpret the size of a tracer particle using the Stokes-Einstein relation, $D_0 = \frac{k_B T}{6\pi \eta a}$. For a particle in water at $T = 300 \, \text{K}$ (where $\eta \approx 10^{-3} \, \text{Pa}\cdot\text{s}$), this diffusion coefficient corresponds to a hydrodynamic radius of $a \approx 0.22 \, \mu\text{m}$. Moreover, we can use these parameters to calculate the amplitude $\delta A_0$ of channel wall undulations using Eq.~\ref{eq:63}, which yields $\sim 0.6 \;\mu$m.

\begin{figure}[ht]
\includegraphics[width=8.5
cm,height=6.0cm]{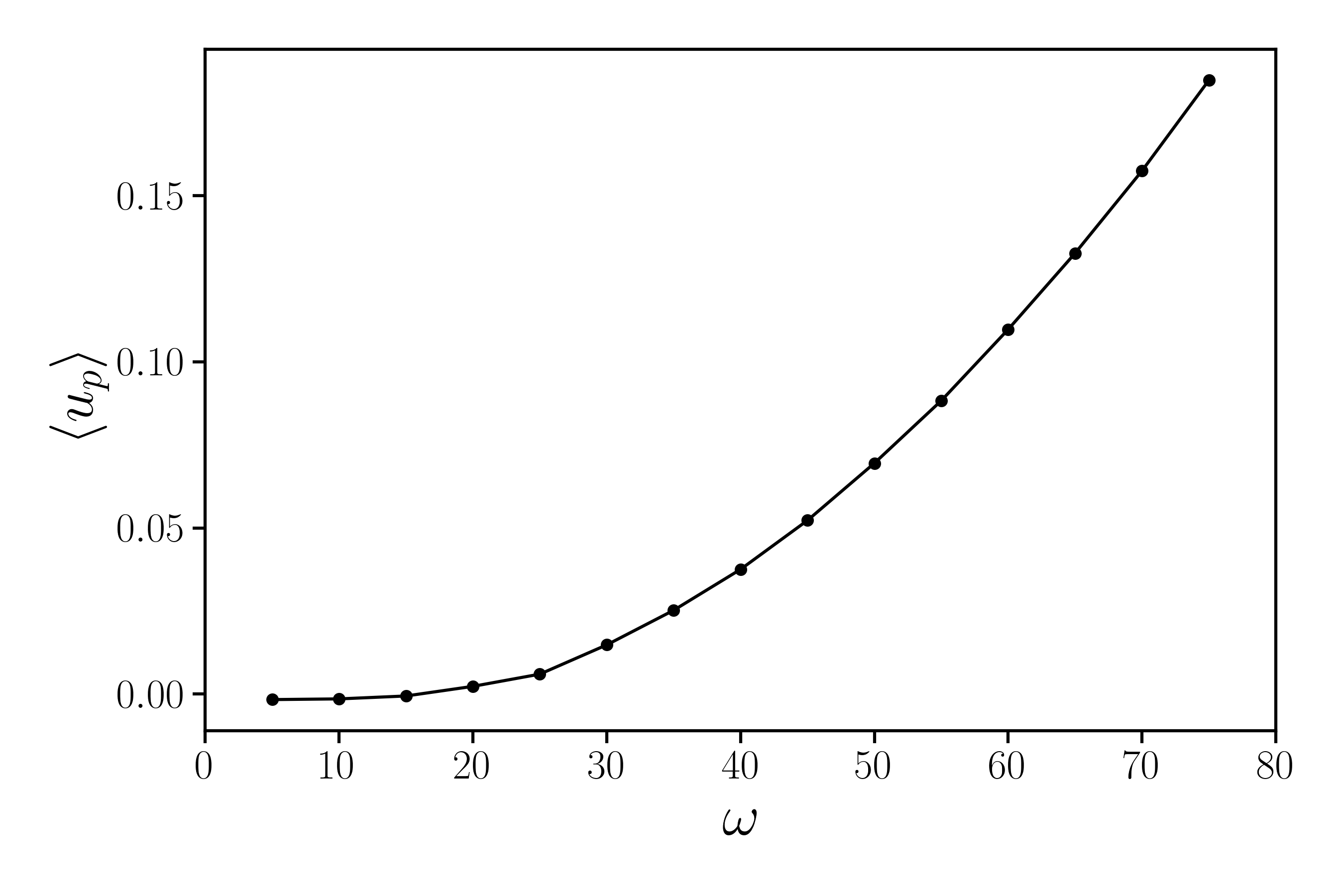}
\caption{Variation of average axial velocity with the frequency of wall undulations ($\langle u_p \rangle$ is in $\mu\text{m s}^{-1}$ and $\omega$ is in $\text{rad s}^{-1}$). The simulation parameters used are: $R_0 = 10~\mu\text{m},~ v_{\text{slip}} = 1.0~\mu\text{m s}^{-1},~ \lambda = 1.0~\mu\text{m},~ \phi = -\pi/2,~ D_0 = 1.0~\mu\text{m}^2\text{ s}^{-1}$.}

\label{fig:3}       
\end{figure}

\begin{figure}[ht]
\includegraphics[width=8.5
cm,height=6.0cm]{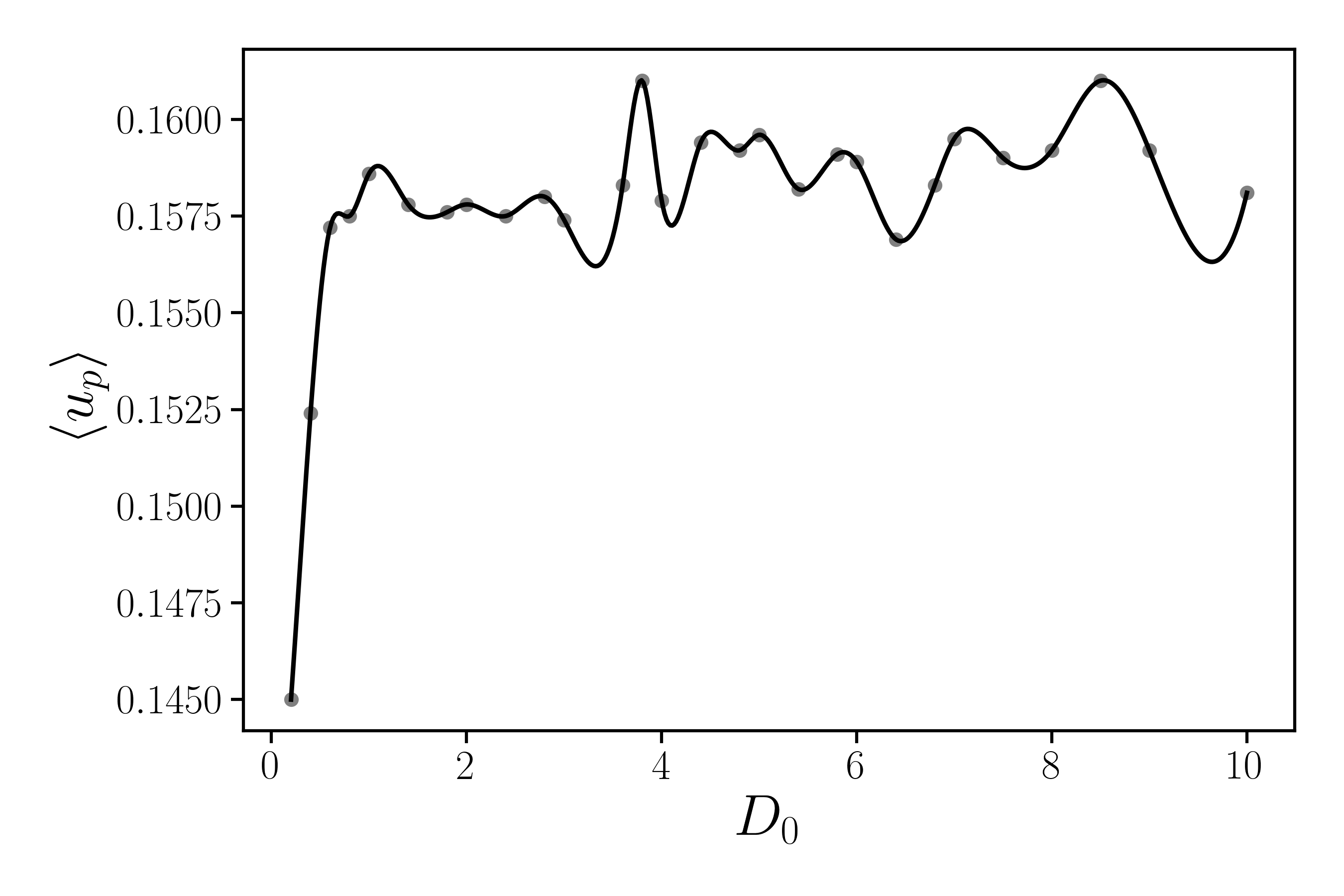}
\caption{Variation of average axial velocity with the diffusivity $D_0$ of the tracer particle ($D_0$ is in $\mu\text{m}^2\text{ s}^{-1}$ and $\langle u_p \rangle$ is in $\mu\text{m s}^{-1}$). The simulation parameters used are: $R_0 = 10~\mu\text{m},~ v_{\text{slip}} = 1.0~\mu\text{m s}^{-1},~ \lambda = 1.0~\mu\text{m},~ \phi = -\pi/2,~ \omega = 70~\text{rad s}^{-1}$.}

\label{fig:4}       
\end{figure}

Next, we investigate the variation of the ratcheting speed with the external forcing parameter, specifically the frequency $\omega$ to the square of which is proportional to the non-equilibrium energy input to the ratcheting tracer particle. In Fig.~\ref{fig:3}, we show the variation of ratcheting speed $\langle u_p \rangle$ (mean velocity along the axis of the channel) with the frequency $\omega$ of the surface undulations. The simulation parameters used are: $R_0 = 10~\mu\text{m},~ v_{\text{slip}} = 1.0~\mu\text{m s}^{-1},~ \lambda = 1.0~\mu\text{m},~ \phi = -\pi/2,~ D_0 = 1.0~\mu\text{m}^2\text{ s}^{-1}$. We find that the ratcheting speed (mean axial velocity) increases with increasing undulation frequency. 
In this analysis, the angular frequency $\omega$ is varied from 0 to 75 rad/sec. We chose this range because exceeding this limit would cause the $\omega$ dependent terms, specifically the surface undulation amplitude and the first-order velocity correction $u_1$ in Eq.~\ref{eq:31} and \ref{eq:63} to become too large, violating the assumptions of the perturbation method. 

Here, a note on the condition for ratcheting is in order. The condition of ratcheting is that the diffusive length of the tracer particle, that is, the distance covered by the tracer particle in one time period of oscillation, should be of the order of the wavelength of undulations. This condition on the parameters $\omega$, $k$, and $D_0$ translates into $\omega \sim k^2 D_0$, where the exact formula depends on the details of the symmetry-broken waveform. The $\omega$ and $\lambda$ follow an inverse relationship for the ratcheting to happen effectively, and one of them is a free parameter for a given $D_0$. While setting the ratcheting parameters, we have been guided by this simple principle.

In Fig.~\ref{fig:4}, we show the variation of axial ratcheting speed $\langle u_p \rangle$ with diffusivity $D_0$ of the tracer particle under the condition that the driving field obeys the relation $\omega \sim k^2 D_0$. We find that the ratcheting first increases, then the ratcheting speed becomes practically constant over a large range of diffusivity values. This simply means that a minimum level of diffusion is necessary for a given range of wavelength of undulating modes, and beyond that, a substantial range exists in which the ratcheting remains stable. We have not increased $D_0$ beyond the range shown due to the technical issue that arises with a corresponding increase in $\omega$ for ratcheting to occur. A very high value of $\omega$ will require us to reduce the simulation time steps, slowing the simulation. The simulation parameters used are: $R_0 = 10~\mu\text{m},~ v_{\text{slip}} = 1.0~\mu\text{m s}^{-1},~ \lambda = 1.0~\mu\text{m},~ \phi = -\pi/2,~ \omega = 70~\text{rad s}^{-1}$. 

At this point, we would like to conclude this section by noting that although the present analysis was explicitly conducted for water, the results can be applied to liquids of different natures, including those of biological relevance, such as alcohol, blood plasma, cytoplasm, and many others. For the fixed diffusivity tracer particle at the same room temperature, which is $300\,\mathrm{K}$, a different liquid, such as blood plasma (which is more viscous than water), will allow the same analysis to hold for the transport of smaller particles. 

\section{DISCUSSION}
\par In the present paper, we have derived the velocity field for a spatiotemporally undulating microchannel to construct the equation of motion of a tracer particle by considering the dominant forces in the low Reynolds number regime: the viscous drag and the stochastic force. This approach, known as the quasi-steady Stokesian or Langevin model, is a common and powerful simplification. A complete treatment of particle dynamics in an unsteady low Reynolds number flow is described by the Basset-Boussinesq-Oseen equation (BBO) \cite{maxey1983equation, landau1987fluid, basset1888treatise}. This equation includes additional terms, such as the Froude-Krylov force (due to the pressure gradient), the added mass (the inertia of the surrounding fluid), and the Basset history force (viscous memory of the flow), all of which arise from the flow's unsteady nature.

\par In the steady flow regime, we have shown that a local inversion symmetry broken hydrodynamic field could be modeled in a very systematic manner for a fluid in an undulating microchannel. We demonstrated that a self-consistent relationship exists between surface undulations and fluid velocity profile. We have then shown that a physically meaningful regime of tracer particle ratcheting exists in such hydrodynamic fields, where the particles are dragged by the fluid's velocity. In the absence of the phenomenon of ratcheting, such undulating velocity fields will not result in directional transport of tracers. Due to the decoupling of pressure-driven and surface-driven undulating velocities, the present analysis may be particularly helpful in applications such as filtration. Moreover, the perturbatively accessible analysis in the low-Reynolds-number regime reveals the details of the process's dependence on the fluid and forcing parameters. 

\par The present model assumes the presence of an inversion symmetry broken stationary wave flow profile in the bulk and then, from the fluid dynamic consistency, evaluates the corresponding stationary wave undulations on the boundary, which can excite such bulk flows. It turns out that there are four stationary wave excitations that exist on the boundary as shown in Eq.~\ref{eq:64}. A part of the oscillation frequency of these excitations is a function of the $v_{\rm slip}$ selected by the system. This implies that even if one tries to create such stationary wave modes on, for example, an elastic surface that confines the fluid by driving traveling waves from the ends, it will depend on the system and the selection of $v_{\rm slip}$. This is a practical limitation that one needs to consider in such a context. Otherwise, controlled boundary motions are experimentally feasible in soft microfluidic systems using piezoelectric, acoustic, or electromechanical actuation. In particular, traveling and standing wall deformation waveforms have been realized in PDMS-based microchannels using piezoelectric actuation, where the spatial structure of the deformation is determined by actuator geometry, membrane design, and spatiotemporal phasing of the driving signal \cite{nakahara2011peristaltic, hilber2016stimulus, xu2023compact, friend2011microscale, brandi2024numerical}.

\par Here, it is worth noting that Eq.~\ref{eq:31} reveals a sharp dependence of the undulating velocity amplitude on the channel width as $R_0^3$ and the oscillation frequency as $\omega^2$. Specifically, the cubic and quadratic dependences on $R_0$ and $\omega$ determine the energy input to the ratcheting particle. It renders ratcheting particularly pronounced in channels of micrometer width, thereby making transport very small or impossible in very small-width nanochannels, at least within the purview of the present perturbative analysis. This particular observation could be crucial for biological systems.  

\par
The chosen slip velocity ($v_{\mathrm{slip}} = 1.0~\mu\mathrm{m\,s^{-1}}$) is consistent with values inferred from microfluidic experiments on hydrophobic and polymer-coated channels. Direct micro-PIV measurements by Tretheway and Meinhart \cite{tretheway2002apparent} report a finite wall-adjacent slip velocity of order $10\%$ of the free-stream velocity, corresponding to effective slip lengths of order $\sim 1~\mu\mathrm{m}$. For typical pressure-driven microflows with mean velocities of $10$--$100~\mu\mathrm{m\,s^{-1}}$, this implies slip velocities in the range $0.5$--$10~\mu\mathrm{m\,s^{-1}}$, making the present choice physically reasonable rather than arbitrary. However, it is essential to note that in many biological systems, such flows occur with negligible slip velocity, and the present model, which requires a finite slip velocity, is not applicable to these systems.

\par Another key assumption in our model is the neglect of particle-boundary hydrodynamic interactions (also known as wall-induced hydrodynamic interactions). These interactions depend strongly on the particle's proximity to the boundary and its size relative to the channel dimensions, leading to a position-dependent particle diffusivity. These wall-induced Hydrodynamic interactions become important for the dynamics of a tracer particle in small-scale channels \cite{marbach2018transport}, such as micro- and nanochannels. Taking these effects into account systematically is a non-trivial extension of the model, which will be pursued in future works. 

\par A natural extension of the present work might also consider the dynamics of many interacting tracer particles or structured particles in such a steady inversion symmetry broken velocity field. How mutual cooperativity, specifically interparticle interactions, changes transport properties can be investigated within the present theoretical framework, provided the particles remain tracer particles. The success of the perturbative scheme developed in this paper in the low Reynolds number regime also opens up the possibility of future investigation of ratcheting under various other boundary condition scenarios applied to the tracer.

\section{ACKNOWLEDGEMENTS}
Aakash Anand would like to thank the Council of Scientific and Industrial Research (CSIR), India, for funding this research through Grant No. 09/936(0296)/2021-EMR-I. We thank the National Supercomputing Mission for providing computing resources of ``PARAM Brahma” at IISER Pune, implemented by C-DAC and supported by the Ministry of Electronics and Information Technology (MeitY) and the Department of Science and Technology (DST), Government of India.

\section{DATA AVAILABILITY}

Data supporting the findings of this study are available from the corresponding author upon reasonable request.

\section{DECLARATION OF INTERESTS}
The authors declare that they have no conflict of interest.

\bibliographystyle{unsrt}

\bibliography{references.bib}

\end{document}